\documentclass[aps,prl,superscriptaddress,reprint]{revtex4-1}

\usepackage[usenames, dvipsnames]{color}

\usepackage{graphicx}
\usepackage{xcolor}
\usepackage{rotating}
\usepackage{bm,amsmath,amssymb}
\usepackage[utf8]{inputenc}
\usepackage{footmisc}

\DeclareMathAlphabet{\mathpzc}{OT1}{pzc}{m}{it}

\usepackage{setspace}

\usepackage{slashed}
\usepackage{upgreek}

\newcommand{\nb}[1]{\color{blue}}

\usepackage[
	  pagebackref=false,
	  colorlinks=true,
      linkcolor=blue,
      urlcolor=blue,
      filecolor=black,
      citecolor=red,
      pdfstartview=Fit,
      pdftitle={Toward a Quantum Information Foundation for Neutron Scattering},
        pdfauthor={Michael J. Landry and Mingda Li},
        pdfsubject={},
        pdfkeywords={},
        pdfpagemode=UseNone,
        bookmarksopen=true
      ]{hyperref}

\usepackage{graphicx}
\graphicspath{ {./images/} }

\usepackage{comment}

\usepackage[normalem]{ulem}
\usepackage{amsmath}
\usepackage{enumerate}
\usepackage{amsfonts}
\usepackage{epsfig}
\usepackage{mathbbol}

\def\Tr{\mathop{\rm Tr}}
\def\tr{\mathop{\rm tr}}

\def\Re{\mathop{\rm Re} }

\newcommand\half{{\ensuremath{\frac{1}{2}}}}

\newcommand\vev[1]{{\ensuremath{\left\langle{#1}\right\rangle}}}

\newcommand\ket[1]{\ensuremath{\lvert{#1}\rangle}}
\newcommand\bra[1]{\ensuremath{\langle{#1}\rvert}}

\newcommand{\be}{\begin{equation}}
\newcommand{\ee}{\end{equation}}
\newcommand{\bea}{\begin{eqnarray}}
\newcommand{\eea}{\end{eqnarray}}
\newcommand{\bega}{\begin{gather}}
\newcommand{\eega}{\end{gather}}

\newcommand{\bi}{\begin{itemize}}
\newcommand{\ei}{\end{itemize}}
\newcommand{\ben}{\begin{enumerate}}
\newcommand{\een}{\end{enumerate}}
\newcommand{\bca}{\begin{cases}}
\newcommand{\eca}{\end{cases}}
\newcommand{\bln}{\begin{align}}
\newcommand{\eln}{\end{align}}
\newcommand{\bst}{\begin{split}}
\newcommand{\est}{\end{split}}
\def\ie{\begin{equation}\begin{aligned}}
\def\fe{\end{aligned}\end{equation}}
\newcommand{\bma}{\le(\begin{matrix}}
\newcommand{\ema}{\end{matrix}\ri)}
\newcommand{\bwt}{\begin{widetext}}
\newcommand{\ewt}{\end{widetext}}

\newcommand\Lam{\Lambda}

\newcommand\ha{{\half}}

\def\le{\left}
\def\ri{\right}

\newcommand\sH{{\ensuremath{{\mathcal H}}}}

\newcommand\sS{{\mathcal S}}

\DeclareMathAlphabet{\pazocal}{OMS}{zplm}{m}{n}

\def\doibase#1{\if#1 \expandafter\doibase\else https://doi.org/#1\fi}

\begin{document}

\title{Toward a Quantum Information Foundation for Neutron Scattering}

\author{ Michael J. Landry}
\email{mjlandry@mit.edu}
\affiliation{Quantum Measurement Group, MIT, Cambridge, MA 02139, USA}
\affiliation{Department of Nuclear Science and Engineering, MIT, Cambridge, MA 02139, USA}
\affiliation{Department of Physics, MIT, Cambridge, MA 02139, USA}

\author{Yongqiang Cheng}
\affiliation{Neutron Scattering Division, Oak Ridge National Laboratory, Oak Ridge, TN 37831, USA}

\author{Daniel Pajerowski}
\affiliation{Neutron Scattering Division, Oak Ridge National Laboratory, Oak Ridge, TN 37831, USA}

\author{Joshua J. Turner}
\affiliation{SLAC National Accelerator Laboratory, Menlo Park, CA 94025, USA}

\author{Mingda Li}
\email{mingda@mit.edu}
\affiliation{Quantum Measurement Group, MIT, Cambridge, MA 02139, USA}
\affiliation{Department of Nuclear Science and Engineering, MIT, Cambridge, MA 02139, USA}

\begin{abstract}
Neutron scattering is a central probe of quantum materials, yet what quantum information can, in principle, be extracted from a scattering experiment remains unclear. Here we formulate neutron scattering as a quantum measurement, with the scattering amplitudes acting as Kraus operators on the material state. This provides a unified description of conventional intensity, coherent scattering, interferometric measurements, and entanglement witnesses, including those based on quantum Fisher information. We then ask how far this connection can be extended to nonlinear properties of a quantum state. For two independently prepared samples, we derive the general structure of probe-mediated measurement probabilities and show that, for spin-projection-conserving dynamics, low-order scattering cannot resolve certain many-body coherences required to determine purity. Conversely, an idealized single-spin protocol shows how probe-mediated exchange can access purity when the appropriate two-copy operation is realized. Our results establish a quantum-information description of neutron scattering and delineate the information accessible through conventional and interferometric measurements.
\end{abstract}

\maketitle

Neutron scattering is one of the most incisive probes of microscopic structure and dynamics in condensed matter and materials physics. It resolves nuclear and magnetic correlations across a wide range of systems, including energy materials and ceramics \cite{ramirez2009neutron,niedziela2019selective,sun2022mutual,sun2023spin}, polymers and biological macromolecules \cite{chu2001small,ashkar2018neutron}, superconductors \cite{shirane1969lattice,tranquada2020cuprate,mcqueeney1999anomalous,ma2017prominent,yildirim2001giant}, frustrated magnets \cite{ross2011quantum,stone2006quasiparticle,banerjee2018excitations,han2012fractionalized}, and topological materials \cite{gaudet2021weyl,drucker2023topology,bhattacharjee2022interface,nguyen2020topological}. Its central observables are scattering intensities, resolved in momentum, energy, and polarization, which are related to the dynamic structure factor \cite{lovesey1984,squires2012introduction}. For a fixed measurement setting, the recorded probability has the form $\tr(\hat E\rho)$ and is therefore linear in the sample density matrix $\rho$. This simple observation raises a broader question: what quantum information about a many-body state is actually accessible through scattering?

That question has become increasingly important given the close connection of quantum materials and entanglement. The language of quantum information now plays an important role in the characterization of quantum materials, especially when one seeks to distinguish classical correlations, entanglement, coherence, and mixedness. In scattering experiments, these notions usually enter indirectly, through correlation functions, response functions, or model-based reconstruction. Our aim here is to cast neutron scattering itself in a quantum-information language. We show that the basic scattering amplitude is naturally a Kraus operator, so that neutron scattering may be viewed as a quantum instrument acting on the material state. In this form, conventional intensity, coherent amplitude, and interferometric channel coherence within a single quantum-measurement description.

This viewpoint also sharpens the distinction between different notions of entanglement characterization. A conventional entanglement witness is a Hermitian operator, whose expectation value is nonnegative on every separable state, so a negative value certifies entanglement. Structure-factor witnesses connect this criterion directly to scattering-accessible correlators \cite{krammer2009multipartite}. Entanglement monotones serve a different purpose: they quantify nonclassical resources that cannot increase on average under local operations and classical communication \cite{vidal2000entanglement}. Quantities such as concurrence and tangle are generally nonlinear in the state \cite{wootters1998entanglement}. Their extraction from neutron data therefore requires more than witnessing alone; one must also know which observables have been accessed and what additional assumptions link them to the desired quantity. At finite temperature, this distinction becomes especially important because reduced-state mixedness contains both entanglement and thermal contributions \cite{scheie2025tutorial}.

Quantum Fisher information (QFI) provides a particularly important bridge between spectroscopy and multipartite entanglement. For a chosen generator, surpassing the corresponding $k$-producible bound certifies entanglement depth of at least $k+1$ constituents \cite{hyllus2012fisher}. In thermal equilibrium, the QFI of a Hermitian observable $\hat A$ is related to the absorptive dynamical susceptibility by \cite{hauke2016measuring}
\be\label{qfiThermal}
 F_Q[\rho_\beta,\hat A]=\frac{4}{\pi}\int_0^\infty d\omega\,
 \tanh\!\left(\frac{\beta\omega}{2}\right)\chi_A''(\omega),
\ee
where $\rho_\beta=e^{-\beta\hat H_S}/Z$ and $\hbar=k_B=1$. Through the fluctuation--dissipation relation, neutron spectroscopy therefore gives experimental access to an entanglement witness built from a dynamical response function. This approach has enabled entanglement certification in KCuF$_3$ and Cs$_2$CoCl$_4$, alongside analyses of one- and two-tangle \cite{scheie2021witnessing,laurell2021quantifying,scheie2025tutorial}. At the same time, QFI remains nonlinear in a general density matrix; Eq.~\eqref{qfiThermal} relies on equilibrium and a known temperature. Nor is QFI for a fixed generator, by itself, a general entanglement monotone, although QFI-matrix bounds can constrain such monotones \cite{du2026uncertainty}.

Purity, $\tr\rho^2$, poses a related but distinct problem. It measures mixedness rather than entanglement, although reduced-state purity diagnoses bipartite entanglement when the global state is pure \cite{nielsen2010quantum}. Unlike a single-copy expectation value, purity is quadratic in $\rho$ and is naturally associated with a two-copy swap observable \cite{ekert2002direct}. Direct access to such quantities is well known in optics \cite{hong1987measurement} and in cold atoms \cite{islam2015measuring,kaufman2016quantum}. Neutron interferometry likewise permits coherent control of path and spin \cite{rauch2015neutron,hasegawa2003violation}, and neutron beams mode-entangled in spin, path, and energy have been realized and applied as probes of entangled targets \cite{shen2020unveiling,kuhn2021neutron,irfan2021quantum}, but coherent manipulation of the probes does not automatically imply access to the corresponding two-copy operator on the material state. The issue is therefore not merely nonlinearity, but which operators neutron scattering can realize or estimate.

\begin{figure}[t]
\centering
\includegraphics[width=\columnwidth]{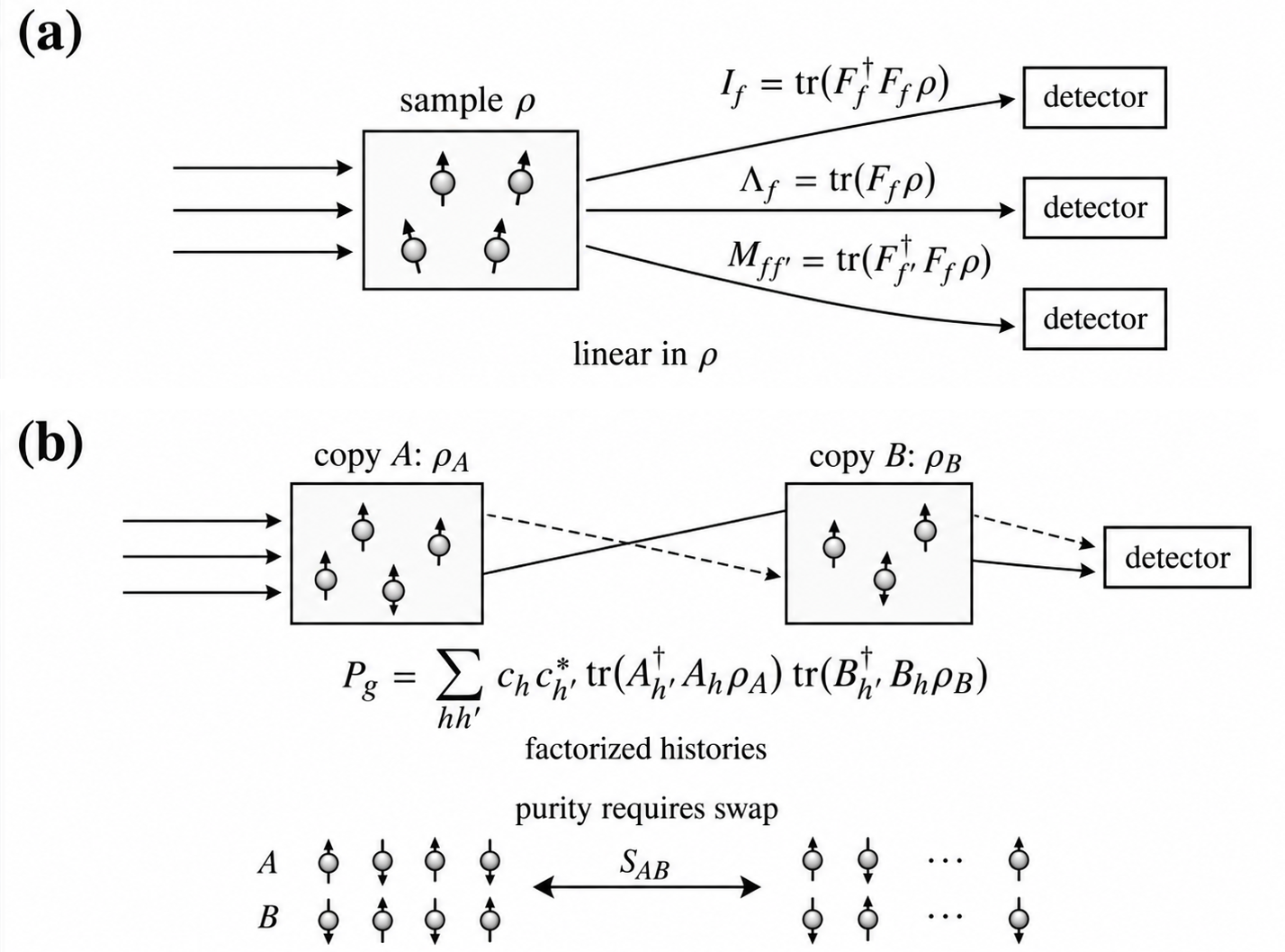}
\caption{\label{fig:schematic}
Quantum-information view of neutron scattering.
(a) For a single sample, neutron scattering measures quantities that are linear in the sample density matrix $\rho$. Conventional intensity, coherent amplitude, and interferometric channel coherence correspond to different combinations of the scattering operators $\hat F_f$.
(b) For two independently prepared copies, different scattering histories can interfere, but the resulting detection probability still separates into contributions associated with each copy. Measuring the purity $\tr(\rho^2)$ requires access to a swap operation $S_{AB}$ that exchanges the states of the two copies.}
\end{figure}

The quantum-instrument viewpoint developed in this Letter provides a common foundation for these questions. It identifies the single-sample observables naturally accessible to scattering, clarifies how interferometry enlarges that set, and organizes the connection between scattering and entanglement witnesses such as QFI. It also gives a clean language for asking when two-copy quantities can be accessed. For two independently prepared samples with no direct coupling, we derive a history-resolved factorization of arbitrary probe-mediated detection probabilities. Although such factorization cannot by itself rule out nonlinear state estimation, it makes clear what additional operator structure is needed. In particular, for spin-projection-conserving dynamics, sufficiently low-order scattering cannot distinguish a coherent many-body superposition from its incoherent mixture, whereas an idealized single-spin exchange protocol shows how purity can emerge when the required two-copy operator is available. Figure~\ref{fig:schematic} summarizes this perspective.

\textit{Scattering as a quantum instrument.---}Consider a neutron incident in channel $i=(\bm k_i,\sigma_i)$ on a sample with density matrix $\rho$ on a Hilbert space $\sH$. The Hamiltonian is $\hat H=\hat H_0+\hat H_S+\hat V$, where $\hat H_0$ describes the free neutron, $\hat H_S$ the sample, and $\hat V$ their interaction. For normalized incident and detected wave-packet channels, the exact conditional operator is $\hat F_f=\bra f\hat U\ket i$, with $\hat U$ the joint evolution. In the Born approximation, its scattering contribution is proportional to the interaction matrix element
\be\label{Fdef}
 \hat F_f\ \propto\ \bra{\bm k_f\sigma_f}\hat V\ket{\bm k_i\sigma_i},
\ee
with the energy filter and kinematic normalization understood. This matrix element acts on the sample Hilbert space. For magnetic scattering, the vertex is linear in the Fourier spin density at $\bm Q=\bm k_i-\bm k_f$, projected perpendicular to $\bm Q$ and contracted with the neutron spin matrix element; the nuclear vertex involves the scattering-length-weighted nuclear density. The energy transfer is $\omega=E_i-E_f$. The operator formulation below is exact; the spin-linear vertex is an additional property used in the bounded-order argument. In a sample energy basis, the energy-resolved amplitude connects states whose energies differ by $\omega$. Unobserved final sample states are summed over in the detection probability. Higher-order scattering replaces the single-vertex amplitude by products of interaction operators and propagators, while retaining the same conditional-measurement structure. Thus, the distinction between a linear measurement probability and a low-order spin correlator must be maintained: the former is general, whereas the latter depends on the scattering approximation.

Detection in channel $f$ applies the unnormalized map $\rho\mapsto\hat F_f\rho\hat F_f^\dagger$. Before scattering, the joint state is $\ket i\bra i\otimes\rho$. Projecting the evolved state onto $\ket f$ leaves the sample block $\hat F_f\rho\hat F_f^\dagger$; the trace of this block is the detection probability. The probe need not resolve the final sample state, which is traced over rather than measured individually. Writing $\tr$ for the sample trace, its probability is
\be\label{If}
 P_f=\tr(\hat F_f\rho\hat F_f^\dagger)
     =\tr(\hat F_f^\dagger\hat F_f\rho)\equiv I_f,
\ee
and the conditional sample state is $\hat F_f\rho\hat F_f^\dagger/I_f$. Thus, $\hat F_f$ is a Kraus operator and $\hat E_f=\hat F_f^\dagger\hat F_f$ the associated positive operator-valued measurement \cite{nielsen2010quantum}. Including all outcomes, including unscattered neutrons, gives the complete instrument. For an exhaustive set of normalized channels, the effects satisfy $\sum_f\hat E_f=\mathbb 1$, with sums replaced by integrals where appropriate. A detector bin that combines unresolved probe channels instead has a sum of Kraus terms. These distinctions affect normalization and resolution, but not linearity in the initial sample state. In the usual scattering limit, $I_f$ is proportional to the appropriate dynamic structure factor, with polarization and kinematic factors. The sample's conditional evolution can depend nonlinearly on $\rho$ through normalization, but every unnormalized outcome probability remains linear.

Interferometry accesses matrix elements beyond the channel intensities. For an incident superposition $a\ket1+b\ket2$, the conditional operator is $a\hat F_{f\leftarrow1}+b\hat F_{f\leftarrow2}$, giving
\be\label{interf}
\begin{aligned}
 P_f={}&|a|^2I_{f\leftarrow1}+|b|^2I_{f\leftarrow2}\\
 &+2\Re\!\left[ab^*\tr(\hat F_{f\leftarrow2}^\dagger
                         \hat F_{f\leftarrow1}\rho)\right].
\end{aligned}
\ee
The interference term is the overlap between the sample amplitudes associated with the two histories. Orthogonal final sample supports eliminate this term, even if the neutron paths are recombined. For a pure initial sample, the term is the inner product of the two conditional sample vectors. For a mixed sample, it is the corresponding Hilbert--Schmidt inner product of $\hat F_{f\leftarrow1}\sqrt\rho$ and $\hat F_{f\leftarrow2}\sqrt\rho$. Consequently, a channel intensity fixes the magnitude of the amplitude but does not determine either the reference phase or the relative overlap of two histories. Superposed neutron spin states yield polarized-neutron interference terms of the Blume--Maleev cross section \cite{blume1963polarization,maleev1963scattering}. A matched reference history that acts as the identity on the sample gives $\tr(\hat F_f\rho)$, allowing access to the coherent amplitude relative to the reference phase \cite{rauch2015neutron}. Coherent recombination of different momentum channels instead probes their mutual coherence, provided the outgoing probe modes are made indistinguishable. These cases are summarized by
\be\label{functionals}
\begin{aligned}
 I_f&=\tr(\hat F_f^\dagger\hat F_f\rho),\qquad
 \Lam_f=\tr(\hat F_f\rho),\\
 M_{ff'}&=\tr(\hat F_{f'}^\dagger\hat F_f\rho).
\end{aligned}
\ee
The Cauchy--Schwarz inequality gives $|M_{ff'}|^2\leq I_fI_{f'}$, with equality when $\hat F_f\sqrt\rho$ and $\hat F_{f'}\sqrt\rho$ are linearly dependent. This condition is stronger than equality of the normalized conditional density matrices alone. The measured phase is always relative to a coherent reference; no absolute phase of an isolated Kraus operator is observable. Likewise, finite overlap of the outgoing probe modes is necessary but not sufficient for interference if distinguishable sample excitations retain information about the scattering history.

Tracing out the sample gives a neutron output density matrix whose channel elements are $M_{ff'}$, with $I_f$ on the diagonal. The output state is a linear map of $\rho$, and any fixed measurement on the neutron yields a linear functional of that state. Explicitly, the joint outgoing state has channel blocks $\ket f\bra{f'}\otimes\hat F_f\rho\hat F_{f'}^\dagger$; tracing these blocks over the sample gives Eq.~\eqref{functionals}. This expression also incorporates neutron spin and path as parts of the channel label. Preparing entangled probes, such as the spin--path mode-entangled neutron beams of Refs.~\cite{shen2020unveiling,kuhn2021neutron,lu2020operator}, or measuring probe coincidences changes the available instrument, but does not alter the linear dependence on a single initial sample density matrix. In particular, nonlinear quantities may still be inferred when sufficient observables or additional state constraints are available, as in the thermal QFI relation, Eq.~\eqref{qfiThermal}.

\textit{Two-copy observables and purity.---}No fixed single-copy operator $\hat O$ satisfies $\tr(\hat O\rho)=\tr\rho^2$ for every $\rho$. For two copies of samples, however,
\be\label{swap}
 \tr\rho^2=\Tr[(\rho\otimes\rho)\sS],
\ee
where $\Tr$ denotes the two-sample trace and $\sS$ exchanges the copies \cite{ekert2002direct}. The exchange acts as $\sS(\ket\phi\otimes\ket\psi)=\ket\psi\otimes\ket\phi$. A measurement of its symmetric and antisymmetric sectors yields the overlap $\tr(\rho_A\rho_B)$, which gives purity when the copies have the same density matrix. Macroscopic equivalence alone does not ensure this condition. The second R\'enyi entropy is then $S_2=-\ln\tr\rho^2$; when applied to a subsystem of a pure state, it is an entanglement entropy. For $N$ spin-$\ha$ degrees of freedom,
\be\label{swapN}
\begin{aligned}
 \sS&=\prod_{j=1}^N\ha(\mathbb 1+\bm\sigma_j\cdot\bm\sigma_j')\\
    &=2^{-N}\sum_{T\subseteq\{1,\ldots,N\}}
                   \prod_{j\in T}\bm\sigma_j\cdot\bm\sigma_j'.
\end{aligned}
\ee
Equivalently, $\sS=2^{-N}\sum_{P\in\mathcal P_N}P\otimes P$, where $\mathcal P_N$ contains the $4^N$ Pauli strings, including the identity. Hence $\tr\rho^2=2^{-N}\sum_P[\tr(P\rho)]^2$. Purity is therefore a sum of products of linear functionals; such a product structure cannot be a no-go criterion. The experimental requirement is access to the swap expectation, an equivalent estimator, or sufficient information to reconstruct the relevant state. Unlike direct interference between the material copies, exchanging two probe particles does not automatically exchange their samples. Equation~\eqref{swapN} contains correlations of all Pauli weights, from one site to the full system. A restriction to low-order correlators therefore need not determine the missing contributions. The number of terms in this expansion is not a universal lower bound on the number of experimental settings: a collective two-copy measurement can combine them, and randomized measurements provide alternative estimators \cite{elben2018renyi,brydges2019probing}. Any obstruction must follow from restrictions on the accessible operators rather than the size of a chosen expansion.

\textit{Two-sample factorization and a bounded-order limit.---}Consider samples $A$ and $B$ prepared in $\rho_A\otimes\rho_B$, and a fixed initial state of any number of neutrons. The samples have no direct coupling:
$\hat H=\hat H_0+\hat H_A+\hat H_B+\hat V_A+\hat V_B$, where $\hat V_X$ acts on the probes and sample $X$. Probe paths may be split and recombined, and sample order may be coherently controlled. For pure initial and detected probe states $\ket i$ and $\ket g$, the conditional sample operator has the history expansion
\be\label{Kg}
 \hat K_g=\bra g\hat U\ket i=\sum_h c_h\hat A_h\otimes\hat B_h,
\ee
and its probability is
\be\label{nogo}
\begin{aligned}
 P_g=\sum_{h,h'}c_hc_{h'}^*\tr(\hat A_{h'}^\dagger\hat A_h\rho_A)\tr(\hat B_{h'}^\dagger\hat B_h\rho_B).
\end{aligned}
\ee
Here $h$ labels scattering histories, and the sums may include time and continuum integrations. For finite sample Hilbert spaces, a finite product-operator expansion also exists independently of the history representation.

To obtain Eq.~\eqref{Kg}, expand the evolution in a time-ordered interaction series and insert complete probe states between successive vertices. Each matrix element of $\hat V_A$ acts on $A$ alone, and each matrix element of $\hat V_B$ on $B$ alone. A free propagator factorizes into probe, $A$, and $B$ contributions. A history therefore contains an ordered product of operators on each sample and a coefficient determined by the probe amplitudes. Operators on different samples commute, allowing them to be grouped without changing their ordering on either sample. Antisymmetrization permutes probe labels and modifies coefficients, but preserves this sample-wise structure. Evaluating $\Tr[\hat K_g(\rho_A\otimes\rho_B)\hat K_g^\dagger]$ gives Eq.~\eqref{nogo}. The argument holds order by order and for a converged full evolution; mixed probe preparations and unresolved outcomes add corresponding sums. Although the probability is linear in either sample state with the other fixed, it is generally quadratic when $\rho_A=\rho_B=\rho$. Nor does the factorization exclude probe-mediated sample coupling. Each factor in Eq.~\eqref{nogo} has the form of a single-sample history overlap, generalizing Eq.~\eqref{interf}. Those histories can contain repeated scattering and probe higher-order correlations beyond the ordinary single-scattering structure factor. The factorization identifies the relevant observables while leaving open how they are accessed experimentally. In particular, a coherent sum of product Kraus operators can generate an effective operation coupling the two samples.

A restriction on accessible histories yields a stronger statement. Assume each sample Hamiltonian conserves its total $S^z$, and each interaction vertex is linear in sample spin operators, changing sample $S^z$ by at most one unit. Consider a measurement description in which every retained history contains at most $k$ vertices on each sample.

\textit{Bounded-order theorem.} Let $\ket a$ and $\ket b$ be orthonormal eigenstates of sample $S^z$ whose eigenvalues differ by $\Delta$. If $2k<\Delta$, all probabilities in Eq.~\eqref{nogo} are identical for two copies of
\be\label{catstates}
\begin{aligned}
 \rho_{\rm coh}&=\ket\psi\bra\psi,\qquad
 \ket\psi=(\ket a+\ket b)/\sqrt2,\\
 \rho_{\rm mix}&=\ha(\ket a\bra a+\ket b\bra b),
\end{aligned}
\ee
although their purities are $1$ and $\ha$.

\textit{Proof.} Free sample propagation preserves $S^z$. A product $\hat A_{h'}^\dagger\hat A_h$ contains at most $2k$ spin-changing vertices, so its matrix elements between $\ket a$ and $\ket b$ vanish when $2k<\Delta$. Its expectation is therefore the same in $\rho_{\rm coh}$ and $\rho_{\rm mix}$. The same argument applies on $B$, and substitution in Eq.~\eqref{nogo} proves the claim. $\square$ Equivalently, the difference between the states is $\rho_{\rm coh}-\rho_{\rm mix}=\ha(\ket a\bra b+\ket b\bra a)$. Every retained single-sample history overlap is orthogonal to this difference under the trace pairing. Neither combining outcomes nor nonlinear processing of their probabilities can distinguish the states within the stated measurement class, because the complete retained probability distributions coincide.

For $\ket a=\ket{\uparrow\cdots\uparrow}$ and $\ket b=\ket{\downarrow\cdots\downarrow}$, $\Delta=N$. Thus, state-independent purity determination is impossible within this measurement class when $k<N/2$, irrespective of how the retained histories are interfered. The bound concerns the total vertex count per sample in each amplitude. Higher-order terms in the full scattering evolution may distinguish the states; the theorem does not cover those contributions, additional sample control, or symmetry-breaking sample dynamics. Meeting the order threshold is necessary for this example but not sufficient to implement a purity measurement. The result is state-independent in the operational sense that no estimator built only from this restricted data can return the correct purity for every input state. It does not prevent a protocol from identifying the purity of a smaller, specified family, such as states constrained to a known thermal ensemble. The limitation is also distinct from measurement noise: even exact probabilities of all retained outcomes leave the two states unresolved.

In a frozen-spin approximation, the same threshold follows from operator support: interference between histories with at most $k$ local spin vertices produces operators supported on at most $2k$ sites. With interacting sample dynamics, however, Heisenberg-evolved operators can spread over many sites, so vertex count alone does not impose that spatial support bound. The conserved-spin-projection argument avoids this assumption. Experimental suppression of high-order multiple scattering may further limit sensitivity, but its scaling requires a specified geometry and cannot be inferred from factorization alone.

\textit{Single-spin access through mediated exchange.---}An idealized two-copy protocol illustrates why the interaction algebra matters. Let each sample be a single spin-$\ha$, $\hat{\bm S}_X=\bm\sigma_X/2$, with a spin-exchange scattering amplitude proportional to $v\bm\sigma_n\cdot\hat{\bm S}_X$. Here $v$ is an effective dimensionless amplitude, with common propagation and normalization factors absorbed. Select histories with one scattering event at each sample and coherently combine the two orders with equal weight. For the order $A$ followed by $B$, the sample-probe amplitude is $v^2(\bm\sigma_n\cdot\hat{\bm S}_B)(\bm\sigma_n\cdot\hat{\bm S}_A)$. Reversing the order reverses the antisymmetric term in the Pauli product, while leaving its scalar exchange part unchanged. Detecting the probe in its incident spin state and selecting the symmetric path combination therefore isolates that exchange contribution. Using the Pauli identity, their symmetric component is
\be\label{Kplus}
\begin{aligned}
 \hat K_+&=\frac{v^2}{2}\bra{\sigma_i}
 \big\{\bm\sigma_n\cdot\hat{\bm S}_B,
       \bm\sigma_n\cdot\hat{\bm S}_A\big\}\ket{\sigma_i}\\
 &=v^2\hat{\bm S}_A\cdot\hat{\bm S}_B.
\end{aligned}
\ee
This expression assumes matched path amplitudes and isolation of the selected two-event contribution. Since $\hat{\bm S}_A\cdot\hat{\bm S}_B=\sS/2-\mathbb 1/4$, the associated leading-order probability is
\be\label{singlespin}
 P_+=\frac{|v|^4}{16}\left(5-4\tr[\rho_A\rho_B]\right).
\ee
For identical copies, the overlap is the purity. The result is compatible with Eq.~\eqref{nogo} and lies outside the obstruction $2k<\Delta$ for a single spin. The absence of a direct sample--sample interaction is not an obstacle: the common probe supplies the two-sample operator, and the detection probability contains the desired overlap after a known offset and normalization are removed. Realizing Eq.~\eqref{Kplus} requires the assumed exchange form of the coupling and control of the selected histories beyond the generic magnetic scattering alone.

For an extended sample, the analogous second-order term is a form-factor-weighted collective exchange, $\sum_{jl}f_jg_l\hat{\bm S}_j^A\cdot\hat{\bm S}_l^B$, rather than the product of site-paired swaps in Eq.~\eqref{swapN}. This particular construction therefore does not directly measure arbitrary many-body purity. Whether higher-order histories and additional controls generate the required operator depends on the accessible interaction algebra. The single-spin case is an illustration, not a unique exception to a universal prohibition. For one spin, informationally complete polarization measurements also determine purity through $\tr\rho^2=\ha(1+4|\vev{\hat{\bm S}}|^2)$.

\textit{Example: the Heisenberg dimer.---}Consider two spin-$\ha$ moments at the origin and $\ell\hat{\bm e}_z$, with momentum transfer along the bond. A polarization channel selecting the component along $\hat{\bm e}\perp\hat{\bm e}_z$ gives
$\hat F(Q_z)\propto\hat{\bm e}\cdot(\bm\sigma_1+e^{i\ell Q_z}\bm\sigma_2)$.
We consider the equal-time, energy-integrated limit, suppressing common prefactors. The singlet ground state and infinite-temperature state are
\be\label{dimerstates}
\begin{aligned}
 \rho_0&=\ket s\bra s,\qquad
 \ket s=(\ket{\uparrow\downarrow}-\ket{\downarrow\uparrow})/\sqrt2,\\
 \rho_{\rm th}&=\mathbb 1/4,
\end{aligned}
\ee
with purities $1$ and $1/4$. Both have $\vev{\bm\sigma_i}=0$ and hence $\Lam=0$: their mean magnetic scattering amplitudes vanish, although their intensities do not. Using $\vev{\sigma_1^a\sigma_2^b}=-\delta^{ab}$ for the singlet and zero for the thermal mixture gives $I_0=2-2\cos(\ell Q_z)$ and $I_{\rm th}=2$. At equal selected polarization, the channel coherences are
\be\label{dimerM}
\begin{aligned}
 M_0&=(1-e^{i\ell Q_z})(1-e^{-i\ell Q_z'}),\\
 M_{\rm th}&=1+e^{i\ell(Q_z-Q_z')}.
\end{aligned}
\ee

For the singlet, $|M_0|^2=I_0I_0'$: the bound is saturated because $\bm\sigma_2\ket s=-\bm\sigma_1\ket s$, so each nonzero channel prepares the same triplet amplitude up to a scalar. For the thermal mixture,
$|M_{\rm th}|^2=I_{\rm th}I_{\rm th}'\cos^2[\ell(Q_z-Q_z')/2]$.
The reduced visibility reflects the distinguishability of the sample amplitudes associated with different channels. Thus, interferometry accesses off-diagonal channel information that is absent from intensity measurements alone. The entangled-probe scattering theory of Ref.~\cite{irfan2021quantum}, which treats the same dimer, finds the complementary effect: a spin--path entangled neutron sees its Young-like interference erased when the dimer is maximally entangled.

This visibility is not a state-independent purity estimator. Let $\ket\pm_{\hat{\bm e}}$ denote eigenstates of the measured transverse spin component. The separable mixture
$\rho_{\rm cl}=\ha(\ket{+-}\bra{+-}+\ket{-+}\bra{-+})$, with both kets and bras expressed in this basis,
has purity $1/2$ but the same $I$, $\Lam$, and $M$ as the singlet for this component. Measurements of additional components can distinguish the states. For example, $\hat W=(\mathbb 1+\bm\sigma_1\cdot\bm\sigma_2)/4$ is a two-spin entanglement witness: its expectation is nonnegative for separable states and equals $-1/2$ in the singlet. Measuring the required components therefore certifies entanglement even though the single-component visibility does not determine purity. This separates the information needed for a witness from that needed for a nonlinear state functional. An informationally complete set of one- and two-spin correlators would reconstruct a dimer density matrix and hence its purity, but completeness is not guaranteed by a particular polarized-scattering geometry. For larger systems, reconstruction requires correspondingly more information; restricted scattering observables may still certify entanglement without determining the full state.

\textit{Conclusion.---}Neutron scattering admits a natural formulation in the language of quantum information. Viewed as a quantum instrument, it provides a unified description of conventional intensities, coherent amplitudes, and interferometric channel coherences, while making clear that the experimentally recorded probabilities remain linear in the initial sample state. This perspective brings scattering-based entanglement witnesses, including structure-factor criteria and thermally inferred QFI, into the same operational framework.

For two independently prepared samples, the history expansion of probe-mediated scattering yields a factorized structure for detection probabilities. That structure is not, by itself, a no-go statement for nonlinear state estimation; rather, it isolates the operator content that must be supplied to access genuinely two-copy quantities. The bounded-order result derived here shows one concrete limitation: with spin-projection-conserving sample dynamics, low-order scattering cannot universally resolve the many-body coherences that distinguish certain pure states from their incoherent mixtures. At the same time, the single-spin exchange example illustrates how purity can become accessible when an appropriate two-copy operator is effectively generated.

These results place neutron scattering, entanglement certification, and many-body state characterization within a common conceptual foundation. They suggest a constructive program: use the quantum-instrument structure to identify which observables are naturally available in scattering, which can be promoted by interferometric control, and which require genuinely new probe couplings or auxiliary protocols. In this sense, quantum information is not only a language for interpreting neutron data, but also a guide for extending what neutron scattering can reveal about complex quantum matter.

\begin{acknowledgments}
This work is primarily supported by the U.S. Department of Energy, Office of Science, Basic Energy Sciences
(BES) under the Genesis Mission BES AI Pathfinder
Program, MAIQMag: Multimodal AI for 2D Quantum
Magnets. MJL and ML acknowledge the support from the DOE BES Award No. DE-SC0020148, and support from R. Wachnik. 
\end{acknowledgments}

\bibliography{references}

@book{lovesey1984,
  author = {Lovesey, Stephen W.},
  title = {Theory of Neutron Scattering from Condensed Matter},
  volume = {1 and 2},
  publisher = {Clarendon Press},
  address = {Oxford},
  year = {1984}
}

@article{ramirez2009neutron,
  title = {Neutron scattering and hydrogen storage},
journal = {Materials Today},
volume = {12},
number = {11},
pages = {54-61},
year = {2009},
issn = {1369-7021},
doi = {10.1016/S1369-7021(09)70299-8},
url = {https://www.sciencedirect.com/science/article/pii/S1369702109702998},
author = {A.J. Ramirez-Cuesta and M.O. Jones and W.I.F. David}
}

@article{niedziela2019selective,
  title = {Selective breakdown of phonon quasiparticles across superionic transition in {CuCrSe$_2$}},
  author={Niedziela, Jennifer L and Bansal, Dipanshu and May, Andrew F and Ding, Jingxuan and Lanigan-Atkins, Tyson and Ehlers, Georg and Abernathy, Douglas L and Said, Ayman and Delaire, Olivier},
  journal={Nature Physics},
  volume={15},
  number={1},
  pages={73--78},
  year={2019},
  publisher={Nature Publishing Group UK London},
    url={https://doi.org/10.1038/s41567-018-0298-2}
}

@article{chu2001small,
  title = {Small-angle {X}-ray scattering of polymers},
  doi = {10.1021/cr9900376},
  author={Chu, Benjamin and Hsiao, Benjamin S},
  journal={Chemical reviews},
  volume={101},
  number={6},
  pages={1727},
  year={2001},
  publisher={ACS Publications},
URL = {https://pubs.acs.org/doi/10.1021/cr9900376}
}

@article{ashkar2018neutron,
  title = {Neutron scattering in the biological sciences: progress and prospects},
  author={Ashkar, Rana and Bilheux, Hassina and Bordallo, Heliosa and Briber, Robert and Callaway, David JE and Cheng, Xiaolin and Chu, X-Q and Curtis, Joseph E and Dadmun, Mark and Fenimore, Paul and others},
  journal={Acta Crystallographica Section D: Structural Biology},
  volume={74},
  number={12},
  pages={1129--1168},
  year={2018},
  publisher={International Union of Crystallography},
URL = {https://scripts.iucr.org/cgi-bin/paper?s2059798318017503}
}

@article{shirane1969lattice,
  title = {Lattice-dynamical study of the 110 {K} phase transition in {SrTiO$_3$}},
  author = {Shirane, G. and Yamada, Y.},
  journal = {Phys. Rev.},
  volume = {177},
  issue = {2},
  pages = {858--863},
  numpages = {0},
  year = {1969},
  month = {Jan},
  publisher = {American Physical Society},
  doi = {10.1103/PhysRev.177.858},
  url = {https://link.aps.org/doi/10.1103/PhysRev.177.858}
}

@article{tranquada2020cuprate,
  title = {Cuprate superconductors as viewed through a striped lens},
  doi = {10.1080/00018732.2021.1935698},
  author={Tranquada, John M},
  journal={Advances in Physics},
  volume={69},
  number={4},
  pages={437--509},
  year={2020},
  publisher={Taylor \& Francis},
    url={https://doi.org/10.1080/00018732.2021.1935698}
}

@article{mcqueeney1999anomalous,
  title = {Anomalous dispersion of {LO} phonons in {La$_{1.85}$Sr$_{0.15}$CuO$_4$} at low temperatures},
  author = {McQueeney, R. J. and Petrov, Y. and Egami, T. and Yethiraj, M. and Shirane, G. and Endoh, Y.},
  journal = {Phys. Rev. Lett.},
  volume = {82},
  issue = {3},
  pages = {628--631},
  numpages = {0},
  year = {1999},
  month = {Jan},
  publisher = {American Physical Society},
  doi = {10.1103/PhysRevLett.82.628},
  url = {https://link.aps.org/doi/10.1103/PhysRevLett.82.628}
}

@article{ma2017prominent,
  title = {Prominent role of spin-orbit coupling in {FeSe} revealed by inelastic neutron scattering},
  author = {Ma, Mingwei and Bourges, Philippe and Sidis, Yvan and Xu, Yang and Li, Shiyan and Hu, Biaoyan and Li, Jiarui and Wang, Fa and Li, Yuan},
  journal = {Phys. Rev. X},
  volume = {7},
  issue = {2},
  pages = {021025},
  numpages = {7},
  year = {2017},
  month = {May},
  publisher = {American Physical Society},
  doi = {10.1103/PhysRevX.7.021025},
  url = {https://link.aps.org/doi/10.1103/PhysRevX.7.021025}
}

@article{yildirim2001giant,
  title = {Giant anharmonicity and nonlinear electron-phonon coupling in {MgB$_2$}: A combined first-principles calculation and neutron scattering study},
  author = {Yildirim, T. and G\"ulseren, O. and Lynn, J. W. and Brown, C. M. and Udovic, T. J. and Huang, Q. and Rogado, N. and Regan, K. A. and Hayward, M. A. and Slusky, J. S. and He, T. and Haas, M. K. and Khalifah, P. and Inumaru, K. and Cava, R. J.},
  journal = {Phys. Rev. Lett.},
  volume = {87},
  issue = {3},
  pages = {037001},
  numpages = {4},
  year = {2001},
  month = {Jun},
  publisher = {American Physical Society},
  doi = {10.1103/PhysRevLett.87.037001},
  url = {https://link.aps.org/doi/10.1103/PhysRevLett.87.037001}
}

@article{ross2011quantum,
  title = {Quantum Excitations in Quantum Spin Ice},
  author = {Ross, Kate A. and Savary, Lucile and Gaulin, Bruce D. and Balents, Leon},
  journal = {Phys. Rev. X},
  volume = {1},
  issue = {2},
  pages = {021002},
  numpages = {10},
  year = {2011},
  month = {Oct},
  publisher = {American Physical Society},
  doi = {10.1103/PhysRevX.1.021002},
  url = {https://link.aps.org/doi/10.1103/PhysRevX.1.021002}
}

@article{stone2006quasiparticle,
  title = {Quasiparticle breakdown in a quantum spin liquid},
  author={Stone, Matthew B and Zaliznyak, Igor A and Hong, Tao and Broholm, Collin L and Reich, Daniel H},
  journal={Nature},
  volume={440},
  number={7081},
  pages={187--190},
  year={2006},
  publisher={Nature Publishing Group UK London},
url={https://www.nature.com/articles/nature04593}
}

@article{banerjee2018excitations,
  title = {Excitations in the field-induced quantum spin liquid state of $\alpha$-{RuCl$_3$}},
  author={Banerjee, Arnab and Lampen-Kelley, Paula and Knolle, Johannes and Balz, Christian and Aczel, Adam Anthony and Winn, Barry and Liu, Yaohua and Pajerowski, Daniel and Yan, Jiaqiang and Bridges, Craig A and others},
  journal={npj Quantum Materials},
  volume={3},
  number={1},
  pages={8},
  year={2018},
  publisher={Nature Publishing Group UK London},
URL = {https://doi.org/10.1038/s41535-018-0079-2}
}

@article{han2012fractionalized,
  title = {Fractionalized excitations in the spin-liquid state of a kagome-lattice antiferromagnet},
  author={Han, Tian-Heng and Helton, Joel S and Chu, Shaoyan and Nocera, Daniel G and Rodriguez-Rivera, Jose A and Broholm, Collin and Lee, Young S},
  journal={Nature},
  volume={492},
  number={7429},
  pages={406--410},
  year={2012},
  publisher={Nature Publishing Group UK London},
URL ={https://www.nature.com/articles/nature11659}
}

@article{gaudet2021weyl,
  title = {Weyl-mediated helical magnetism in {NdAlSi}},
  author={Gaudet, Jonathan and Yang, Hung-Yu and Baidya, Santu and Lu, Baozhu and Xu, Guangyong and Zhao, Yang and Rodriguez-Rivera, Jose A and Hoffmann, Christina M and Graf, David E and Torchinsky, Darius H and others},
  journal={Nature materials},
  volume={20},
  number={12},
  pages={1650--1656},
  year={2021},
  publisher={Nature Publishing Group UK London},
url={https://www.nature.com/articles/s41563-021-01062-8}
}

@article{drucker2023topology,
  title = {Topology stabilized fluctuations in a magnetic nodal semimetal},
  author={Drucker, Nathan C and Nguyen, Thanh and Han, Fei and Siriviboon, Phum and Luo, Xi and Andrejevic, Nina and Zhu, Ziming and Bednik, Grigory and Nguyen, Quynh T and Chen, Zhantao and others},
  journal={Nature communications},
  volume={14},
  number={1},
  pages={5182},
  year={2023},
  publisher={Nature Publishing Group UK London},
url={https://www.nature.com/articles/s41467-023-40765-1}
}

@article{bhattacharjee2022interface,
  title = {Interface engineering enabled low temperature growth of magnetic insulator on topological insulator},
  author={Bhattacharjee, Nirjhar and Mahalingam, Krishnamurthy and Will-Cole, Alexandria and Wei, Yuyi and Fedorko, Adrian and Bowers, Cynthia T and Page, Michael and McConney, Michael and Heiman, Don and Sun, Nian Xiang},
  journal={Advanced Materials Interfaces},
  volume={9},
  number={35},
  pages={2201691},
  year={2022},
  publisher={Wiley Online Library},
URL = {https://doi.org/10.1002/admi.202201691}
}

@article{nguyen2020topological,
  title = {Topological singularity induced chiral {Kohn} anomaly in a {Weyl} semimetal},
  author = {Nguyen, Thanh and Han, Fei and Andrejevic, Nina and Pablo-Pedro, Ricardo and Apte, Anuj and Tsurimaki, Yoichiro and Ding, Zhiwei and Zhang, Kunyan and Alatas, Ahmet and Alp, Ercan E. and Chi, Songxue and Fernandez-Baca, Jaime and Matsuda, Masaaki and Tennant, David Alan and Zhao, Yang and Xu, Zhijun and Lynn, Jeffrey W. and Huang, Shengxi and Li, Mingda},
  journal = {Phys. Rev. Lett.},
  volume = {124},
  issue = {23},
  pages = {236401},
  numpages = {7},
  year = {2020},
  month = {Jun},
  publisher = {American Physical Society},
  doi = {10.1103/PhysRevLett.124.236401},
  url = {https://link.aps.org/doi/10.1103/PhysRevLett.124.236401}
}

@article{sun2022mutual,
  title = {Mutual spin-phonon driving effects and phonon eigenvector renormalization in nickel ({II}) oxide},
  author = {Qiyang Sun  and Bin Wei  and Yaokun Su  and Hillary Smith  and Jiao Y. Y. Lin  and Douglas L. Abernathy  and Chen Li },
journal = {Proceedings of the National Academy of Sciences},
volume = {119},
number = {29},
pages = {e2120553119},
year = {2022},
doi = {10.1073/pnas.2120553119},
URL = {https://www.pnas.org/doi/abs/10.1073/pnas.2120553119}
}

@article{sun2023spin,
  title = {Spin-phonon interactions induced anomalous thermal conductivity in nickel ({II}) oxide},
journal = {Materials Today Physics},
volume = {35},
pages = {101094},
year = {2023},
issn = {2542-5293},
doi = {10.1016/j.mtphys.2023.101094},
url = {https://www.sciencedirect.com/science/article/pii/S254252932300130X},
author = {Qiyang Sun and Songrui Hou and Bin Wei and Yaokun Su and Victor Ortiz and Bo Sun and Jiao Y.Y. Lin and Hillary Smith and Sergey Danilkin and Douglas L. Abernathy and Richard Wilson and Chen Li}
}

@article{hong1987measurement,
  title = {Measurement of subpicosecond time intervals between two photons by interference},
  author = {Hong, C. K. and Ou, Z. Y. and Mandel, L.},
  journal = {Phys. Rev. Lett.},
  volume = {59},
  issue = {18},
  pages = {2044--2046},
  year = {1987},
  publisher = {American Physical Society},
  doi = {10.1103/PhysRevLett.59.2044},
  url = {https://link.aps.org/doi/10.1103/PhysRevLett.59.2044}
}

@book{rauch2015neutron,
  title = {Neutron Interferometry: Lessons in Experimental Quantum Mechanics, Wave-Particle Duality, and Entanglement},
  author = {Rauch, Helmut and Werner, Samuel A.},
  edition = {2nd},
  year = {2015},
  publisher = {Oxford University Press},
  address = {Oxford}
}

@article{hasegawa2003violation,
  title = {Violation of a Bell-like inequality in single-neutron interferometry},
  author = {Hasegawa, Yuji and Loidl, Rudolf and Badurek, Gerald and Baron, Matthias and Rauch, Helmut},
  journal = {Nature},
  volume = {425},
  number = {6953},
  pages = {45--48},
  year = {2003},
  publisher = {Nature Publishing Group},
  doi = {10.1038/nature01881}
}

@book{squires2012introduction,
  title = {Introduction to the Theory of Thermal Neutron Scattering},
  author = {Squires, G. L.},
  edition = {3rd},
  year = {2012},
  publisher = {Cambridge University Press},
  address = {Cambridge},
  doi = {10.1017/CBO9781139107808}
}

@article{blume1963polarization,
  title = {Polarization Effects in the Magnetic Elastic Scattering of Slow Neutrons},
  author = {Blume, M.},
  journal = {Phys. Rev.},
  volume = {130},
  issue = {5},
  pages = {1670--1676},
  year = {1963},
  publisher = {American Physical Society},
  doi = {10.1103/PhysRev.130.1670},
  url = {https://link.aps.org/doi/10.1103/PhysRev.130.1670}
}

@article{maleev1963scattering,
  title = {The scattering of slow neutrons by complex magnetic structures},
  author = {Maleev, S. V. and Bar'yakhtar, V. G. and Suris, R. A.},
  journal = {Sov. Phys. Solid State},
  volume = {4},
  pages = {2533},
  year = {1963},
  note = {[Fiz. Tverd. Tela {\bf 4}, 3461 (1962)]}
}

@article{ekert2002direct,
  title = {Direct Estimations of Linear and Nonlinear Functionals of a Quantum State},
  author = {Ekert, Artur K. and Alves, Carolina Moura and Oi, Daniel K. L. and Horodecki, Micha\l{} and Horodecki, Pawe\l{} and Kwek, L. C.},
  journal = {Phys. Rev. Lett.},
  volume = {88},
  issue = {21},
  pages = {217901},
  year = {2002},
  publisher = {American Physical Society},
  doi = {10.1103/PhysRevLett.88.217901},
  url = {https://link.aps.org/doi/10.1103/PhysRevLett.88.217901}
}

@article{islam2015measuring,
  title = {Measuring entanglement entropy in a quantum many-body system},
  author = {Islam, Rajibul and Ma, Ruichao and Preiss, Philipp M. and Tai, M. Eric and Lukin, Alexander and Rispoli, Matthew and Greiner, Markus},
  journal = {Nature},
  volume = {528},
  number = {7580},
  pages = {77--83},
  year = {2015},
  publisher = {Nature Publishing Group},
  doi = {10.1038/nature15750}
}

@article{kaufman2016quantum,
  title = {Quantum thermalization through entanglement in an isolated many-body system},
  author = {Kaufman, Adam M. and Tai, M. Eric and Lukin, Alexander and Rispoli, Matthew and Schittko, Robert and Preiss, Philipp M. and Greiner, Markus},
  journal = {Science},
  volume = {353},
  number = {6301},
  pages = {794--800},
  year = {2016},
  doi = {10.1126/science.aaf6725}
}

@article{brydges2019probing,
  title = {Probing R\'enyi entanglement entropy via randomized measurements},
  author = {Brydges, Tiff and Elben, Andreas and Jurcevic, Petar and Vermersch, Beno\^{\i}t and Maier, Christine and Lanyon, Ben P. and Zoller, Peter and Blatt, Rainer and Roos, Christian F.},
  journal = {Science},
  volume = {364},
  number = {6437},
  pages = {260--263},
  year = {2019},
  doi = {10.1126/science.aau4963}
}

@article{elben2018renyi,
  title = {R\'enyi Entropies from Random Quenches in Atomic Hubbard and Spin Models},
  author = {Elben, A. and Vermersch, B. and Dalmonte, M. and Cirac, J. I. and Zoller, P.},
  journal = {Phys. Rev. Lett.},
  volume = {120},
  issue = {5},
  pages = {050406},
  year = {2018},
  publisher = {American Physical Society},
  doi = {10.1103/PhysRevLett.120.050406}
}

@article{hauke2016measuring,
  title = {Measuring multipartite entanglement through dynamic susceptibilities},
  author = {Hauke, Philipp and Heyl, Markus and Tagliacozzo, Luca and Zoller, Peter},
  journal = {Nature Physics},
  volume = {12},
  number = {8},
  pages = {778--782},
  year = {2016},
  publisher = {Nature Publishing Group},
  doi = {10.1038/nphys3700}
}

@article{scheie2021witnessing,
  title = {Witnessing entanglement in quantum magnets using neutron scattering},
  author = {Scheie, A. and Laurell, Pontus and Samarakoon, A. M. and Lake, B. and Nagler, S. E. and Granroth, G. E. and Okamoto, S. and Alvarez, G. and Tennant, D. A.},
  journal = {Phys. Rev. B},
  volume = {103},
  issue = {22},
  pages = {224434},
  year = {2021},
  publisher = {American Physical Society},
  doi = {10.1103/PhysRevB.103.224434},
  url = {https://link.aps.org/doi/10.1103/PhysRevB.103.224434}
}

@article{laurell2021quantifying,
  title = {Quantifying and Controlling Entanglement in the Quantum Magnet {Cs$_2$CoCl$_4$}},
  author = {Laurell, Pontus and Scheie, Allen and Mukherjee, Chiron J. and Koza, Michael M. and Enderle, Mechtild and Tylczynski, Zbigniew and Okamoto, Satoshi and Coldea, Radu and Tennant, D. Alan and Alvarez, Gonzalo},
  journal = {Phys. Rev. Lett.},
  volume = {127},
  issue = {3},
  pages = {037201},
  year = {2021},
  publisher = {American Physical Society},
  doi = {10.1103/PhysRevLett.127.037201},
  url = {https://link.aps.org/doi/10.1103/PhysRevLett.127.037201}
}

@book{nielsen2010quantum,
  title = {Quantum Computation and Quantum Information},
  author = {Nielsen, Michael A. and Chuang, Isaac L.},
  edition = {10th Anniversary},
  year = {2010},
  publisher = {Cambridge University Press},
  address = {Cambridge}
}

@article{krammer2009multipartite,
  author = {Krammer, Philipp and Kampermann, Hermann and Bru{\ss}, Dagmar and Bertlmann, Reinhold A. and Kwek, Leong Chuang and Macchiavello, Chiara},
  title = {Multipartite Entanglement Detection via Structure Factors},
  journal = {Phys. Rev. Lett.},
  volume = {103},
  pages = {100502},
  year = {2009},
  doi = {10.1103/PhysRevLett.103.100502}
}

@article{vidal2000entanglement,
  author = {Vidal, Guifr{\'e}},
  title = {Entanglement monotones},
  journal = {J. Mod. Opt.},
  volume = {47},
  pages = {355},
  year = {2000}
}

@article{wootters1998entanglement,
  author = {Wootters, William K.},
  title = {Entanglement of Formation of an Arbitrary State of Two Qubits},
  journal = {Phys. Rev. Lett.},
  volume = {80},
  pages = {2245--2248},
  year = {1998},
  doi = {10.1103/PhysRevLett.80.2245}
}

@article{hyllus2012fisher,
  author = {Hyllus, Philipp and Laskowski, Wies{\l}aw and Krischek, Roland and Schwemmer, Christian and Wieczorek, Witlef and Weinfurter, Harald and Pezz{\'e}, Luca and Smerzi, Augusto},
  title = {{Fisher} information and multiparticle entanglement},
  journal = {Phys. Rev. A},
  volume = {85},
  pages = {022321},
  year = {2012},
  doi = {10.1103/PhysRevA.85.022321}
}

@article{scheie2025tutorial,
  author = {Scheie, Allen and Laurell, Pontus and Simeth, Wolfgang and Dagotto, Elbio and Tennant, D. Alan},
  title = {Tutorial: Extracting entanglement signatures from neutron spectroscopy},
  journal = {Materials Today Quantum},
  volume = {5},
  pages = {100020},
  year = {2025}
}

@article{du2026uncertainty,
  author = {Du, Shaowei and Liu, Shuheng and Fadel, Matteo and Vitagliano, Giuseppe and He, Qiongyi},
  title = {Uncertainty Relations between Quantum {Fisher} Information and Entanglement Monotones},
  journal = {Phys. Rev. Lett.},
  volume = {136},
  pages = {110806},
  year = {2026},
  doi = {10.1103/54mc-2yl3}
}

@article{irfan2021quantum,
  title = {Quantum entangled-probe scattering theory},
  author = {Irfan, Abu Ashik Md and Blackstone, Patrick and Pynn, Roger and Ortiz, Gerardo},
  journal = {New J. Phys.},
  volume = {23},
  pages = {083022},
  year = {2021},
  doi = {10.1088/1367-2630/ac12e0}
}

@article{shen2020unveiling,
  title = {Unveiling contextual realities by microscopically entangling a neutron},
  author = {Shen, J. and Kuhn, S. J. and Dalgliesh, R. M. and de Haan, V. O. and Geerits, N. and Irfan, A. A. M. and Li, F. and Lu, S. and Parnell, S. R. and Plomp, J. and van Well, A. A. and Washington, A. and Baxter, D. V. and Ortiz, G. and Snow, W. M. and Pynn, R.},
  journal = {Nat. Commun.},
  volume = {11},
  pages = {930},
  year = {2020},
  doi = {10.1038/s41467-020-14741-y}
}

@article{kuhn2021neutron,
  title = {Neutron-state entanglement with overlapping paths},
  author = {Kuhn, S. J. and McKay, S. and Shen, J. and Geerits, N. and Dalgliesh, R. M. and Dees, E. and Irfan, A. A. M. and Li, F. and Lu, S. and Vangelista, V. and Baxter, D. V. and Ortiz, G. and Parnell, S. R. and Snow, W. M. and Pynn, R.},
  journal = {Phys. Rev. Res.},
  volume = {3},
  pages = {023227},
  year = {2021},
  doi = {10.1103/PhysRevResearch.3.023227}
}

@article{lu2020operator,
  title = {Operator analysis of contextuality-witness measurements for multimode-entangled single-neutron interferometry},
  author = {Lu, Shufan and Irfan, Abu Ashik Md. and Shen, Jiazhou and Kuhn, Steve J. and Snow, W. Michael and Baxter, David V. and Pynn, Roger and Ortiz, Gerardo},
  journal = {Phys. Rev. A},
  volume = {101},
  pages = {042318},
  year = {2020},
  doi = {10.1103/PhysRevA.101.042318}
}
\end{document}